\documentclass[12pt]{article}

\def\beq{\begin{equation}}
\def\eeq{\end{equation}}
\def\barr{\begin{eqnarray}}
\def\beqa{\begin{eqnarray}}
\def\earr{\end{eqnarray}}
\def\eeqa{\end{eqnarray}}
\def\winf{W_{1+\infty}\ }
\def\u1{\widehat{U(1)}}
\def\v{V\,}
\def\w{W\,}
\def\vb{{\overline V}\,}
\def\wb{{\overline W}\,}

\newcommand{\nl}{\nonumber \\}

\begin{document}

\begin{titlepage}

\begin{center}
\hfill  \quad  \\
\vskip 0.5 cm
{\Large \bf Dynamic structure factor of Calogero-Sutherland 
fluids}

\vspace{0.5cm}

Federico~L.~ BOTTESI$^a$ ,\ \ Guillermo~R.~ZEMBA$^{b,c,}$\footnote{
Fellow of Consejo Nacional de Investigaciones Cient\'{\i}ficas y T\'ecnicas, Argentina.}\\
\medskip
{\em $^a$Facultad de Ingenier\'ia, Universidad de Buenos Aires,}\\
{\em  Av. Paseo Col\'on 850,(C1063ACL) Buenos Aires, Argentina}\\
\medskip
{\em $^b$Departamento de F\'{\i}sica Te\'orica, GIyA, Laboratorio Tandar,}\\
{\em  Comisi\'on Nacional de Energ\'{\i}a At\'omica,} \\
{\em Av.Libertador 8250,(C1429BNP) Buenos Aires, Argentina}\\
\medskip
{\em $^c$Facultad de Ingenier\'ia y Ciencias Agrarias,  Pontificia Universidad Cat\'olica Argentina,}\\
{\em  Av. Alicia Moreau de Justo 1500,(C1107AAZ) Buenos Aires, Argentina}\\

\medskip

\end{center}
\vspace{.3cm}
\begin{abstract}
\noindent
The effective extended conformal field theory with symmetry $\winf \times {\overline \winf}$ that describes
the thermodynamic limit of the Calogero-Sutherland model is considered. The dynamic structure factor of the 
chiral component in the repulsive regime is determined  
and compared with the
corresponding one of the free bosonic theory, given that 
both share isomorphic Hilbert spaces but differ in the
time evolution of quantum states.
In either case, a sharp response function peaked at only one 
resonant frequency is found, and the physical implications of this outcome are addressed. 
Furthermore, a detailed comparison between 
this result and the corresponding one 
obtained in the first quantized formulation
of the Calogero-Sutherland
model is provided. Complete agreement in the parameter region in which both
results overlap is found. This outcome provides further support to the equivalence between the first and second quantized
treatments of the Calogero-Sutherland model, in which the computational advantages from integrability in the 
first formulation 
are paralleled by those from the algebraic structure of the second.
\end{abstract}
\vskip 0.5 cm
%Keyords:Effective field theory,bosonization,Conformal field theory
%PACS numbers:  71.30.+h , 05.30.Rt , 64.70.Tg , 02.30.Ik ,02.20.Uw, 11.15.Yc ,71.45.Lr ,73.22.Lp 
\end{titlepage}
\pagenumbering{arabic}
%-2--------------------------------------------- 
The dynamic structure factor (DSF) is a useful theoretical tool designed to probe the 
internal structure of condensed matter systems, of quantum fluids in particular \cite{pines}. 
It describes the response of a system to external density perturbations, such as frequency
sweeps in both space and time. In particular, we will be focusing on 1D fermionic systems
a zero temperature. 
These have been studied with several methods through the last decades of research. 
In this paper, we shall be concerned with the effective field theory (EFT) of this type 
of systems, which describe the low energy, low momenta limit of them \cite{polch}. In particular,
the algebraic bosonization method is based on the $\winf$ algebra \cite{shen} that describes
in a hierarchical structure  
the low lying excitations about the Fermi points of simple, one component fermionic
systems \cite{flsz}. Among them, the Calogero-Sutherland model (CS) stands out for its special properties 
and it has been, therefore, studied in great detail in both first \cite{cs-reviews} and second quantized 
formulations. The first one enjoys the remarkable tools derived from the integrability 
properties of the model, which may be understood as an elaboration of the Bethe ansatz 
method. The second one, in the formulation of the algebraic bosonization, also
enjoys the remarkable tool given by the mathematical structure of the $\winf$ algebra
and the knowledge of its irreducible representations \cite{kac1}, which lead to the classification
of all possible Hilbert spaces for the class of systems considered here.  
Both approaches may be unified conceptually by the notion of the quantum incompressible
fluid \cite{laugh}, conveniently adapted and extended to the domain of CS systems.

In this letter, we calculate the DSF of the CS model using its $\winf$ bosonized EFT, discuss its physical contents, and compare it to previous results obtained within the first
quantized formulation \cite{pustilink}. Our main motivation is to explore the effects 
of the non-linear dispersion relation implied by the EFT of the CS model, that 
describes density waves of the Benjamin-Ono (BO) class 
\cite{benja,ono,boreview} in the quantum realm \cite{aw,abw,bz}. 
Clearly, we do not expect the same results as those obtained in \cite{pustilink}, as
the EFT contains only the low lying excitations of the CS model, but anticipate a 
correspondence between them in the appropriate limits.

The Sutherland model is a theory for non-relativistic fermions in 1D with
a pairwise interaction potential proportional to the inverse square distance in 3D space.
Consider a system of $N$ non-relativistic spinless fermions constrained to move on a 
circumference of perimeter $L$, with 
Hamiltonian \cite{sut} (in units where $\hbar =1$ and $2m=1$, with $m$ beign 
the mass of the particles)
\beq
h_{CS}=\sum_{j=1}^N\ \left( \frac{1}{i} \frac{\partial}{\partial x_j}\
\right)^2\ +\ g\ \frac{\pi^2}{L^2}\ \sum_{i<j}\ \frac{1}{\sin^2
(\pi(x_i-x_j)/L) }\ ,
\label{ham}
\eeq
where $x_i$ ($i=1,\dots,N$) is the coordinate of the $i$-th particle, and $g$ is 
the dimensionless coupling constant. 
The stability of the ground state imposes $g \geq -1/2$, and there are two possible
regimes: attractive ($-1/2 \leq g < 0$) and repulsive ($0 < g$). 
A common reparametrization of the coupling constant in terms of the variable $\xi$ is given by $g=2 \xi ( \xi -1)$, 
so that $\xi \geq 0$ and $0 \leq \xi < 1$ denotes the attractive regime and $1  < \xi $ the repulsive one. 

The $\winf$ EFT of (\ref{ham}) has been presented in \cite{clz}\cite{cfslz} by 
expressing, in the thermodynamic 
limit, the low energy dynamics of fermions around the 1D Fermi surface (two
isolated Fermi points) in terms of relativistic fields.  
This idea is present in all EFT treatments of 1D fermionic
systems and is generally refer to as bosonization. What is special about this
formulation is that it incorporates the natural $U(1)$ symmetry of fermion number
around each Fermi point through its enveloping
algebra, which is identified as $\winf$, by writing all fermionic fields in
a basis of this algebra. Bosonization takes place by replacing the fermionic
fields that satisfy the $\winf$ algebra by bosonic ones that obey exactly the 
same relations. Moreover, these algebraic relations provide a powerful tool
for making computations, which is a major advantage of this method. 
Alternative EFTs descriptions of the CS model may be found in \cite{kaya,khve,amos,poly}.

The EFT of the CS model is an extended conformal field theory \cite{bpz} with symmetry 
$\winf \times {\overline \winf}$ and central charges $(c,{\overline c})=(1,1)$.
Its Hilbert space is isomorphic to that of a free compactified bosonic field,
but with different dynamics provided by a non-linear hamiltonian that defines a
deformed CFT. Usually CFTs are defined on the complex plane. 
However, in many physical applications like in the case of the CS EFT, fields and operators are naturally defined 
on the circumference $(0\le\theta <2\pi)$, {\it i.e.},
on a compact space. Space-time is therefore a {\it cylinder}, formed by the direct product of the compact spatial and the 
unbounded time coordinates.
There is a conformal mapping between these two spaces, which are the {\it cylinder} of radius $R$
$(u=\tau  -iR\theta)$ and the {\it conformal plane} $(z)$ given by
\beq
z=\exp\left(\frac{u}{R}\right)=
\exp\left(\frac{\tau}{R} -i\theta \right) \ ,
\label{confmap}
\eeq
where $\tau$ is the Euclidean time coordinate
and $x$ is periodic with period $L$: $x = R \theta$ with $R=L/(2 \pi )$.

The scale of energy (in natural units with $\hbar = c =1$, where $c$ is the speed of light in the vacuum)
of the CS theory (restoring the mass $m$ of the particles momentarily, for the purposes of comparison) is:
\beq
E_{CS}\ =\ \frac{2{\pi}^2 n_0^2  \xi}{m}\ =\ \frac{ \pi n_0 \xi N}{mR}\ =\ \frac{v}{R}N\ .
\eeq
In terms of the average particle density $n_0 =N/L$, the value of the Fermi (or plasma) velocity is, :
\beq
v\ =\ \frac{\pi n_0\ \xi}{m}\  =\ v_0\ \xi \ ,
\label{vti2}
\eeq
where $\xi=\left(1+\sqrt{1+2g}\right)/2$ is the only free parameter of the EFT,  
defined after (\ref{ham}). We shall focus on the repulsive regime $\xi > 1$ subsequently just for simplicitly 
of the discussion, as extending
our analysis to the attractive regime follows through. Note that $\xi = 1$ corresponds to the free fermion case.

The effective CS hamiltonian on the $(\tau,\theta)$ cylinder is the following operator \cite{bz}: 
\barr
H_{CS} &=&
\left\{ \ E_{CS}\ 
\left[\frac{\sqrt{\xi}}{4}\,\w_0^0
+\frac{1}{N}\,\w_0^1 +
\frac{1}{N^2}\left(\frac{1}{\sqrt{\xi}}\,\w_0^2
-\frac{\sqrt{\xi}}{12}\,\w_0^0 \right.\right.
\right.\nl
&&-\ \left.\left.\left.
\frac{( \xi -1)}{\xi}\,\sum_{\ell=1}^\infty
\,\ell~\w_{-\ell}^0\,\w_\ell^0\right)
\right]+\left(\,W~\leftrightarrow~{\overline W}\,\right)
\right\}~~~.
\label{hcsf2}
\earr
The operators $~\w_{\ell}^m$ in (\ref{hcsf2}) are the 
lowest (in $m=i+1$, where $i$ is the conformal spin \cite{bpz}) 
generators of the infinite dimensional algebra known
as $\winf$ \cite{shen,kac1}. The terms in the $~\w_{\ell}^m$ ($\wb^i_\ell$) operators describe the 
dynamics at the right $(R)$ (left $(L)$) Fermi point, respectively. 
The relevant commutation relations for the $c=1$ CS EFT are:
\barr
\left[\ \w^0_\ell,\w^0_m\ \right] & = &  c\ \xi \ell\ \delta_{\ell+m,0} ~~~,\nl
\left[\ \w^1_\ell, \w^0_m\ \right] & = & -m\ \w^0_{\ell+m} ~~~,\nl
\left[\ \w^1_\ell, \w^1_m\ \right] & = & (\ell-m)\w^1_{\ell+m} + 
\frac{c}{12}\ell(\ell^2-1) \delta_{\ell+m,0}~~~,\nl
\left[\ \w^2_\ell, \w^0_m\ \right] &=& -2m\ \w^1_{\ell+m}~~~,
\label{walg1}\\
\left[\ \w^2_\ell, \w^1_m\ \right] &=& (\ell-2m)\ \w^2_{\ell+m} -
   \frac{1}{6}\left(m^3-m\right) \w^0_{\ell+m}~~~,\nl
\left[\ \w^2_n, \w^2_m\ \right] &=& (2n-2m)\ \w^3_{n+m}
     +{n-m\over 15}\left( 2n^2 +2m^2 -nm-8 \right) \w^1_{n+m}\nonumber\\
     &&\quad +\ c\ {n(n^2-1)(n^2-4)\over 180}\ \delta_{n+m,0}~~~.\nonumber
\earr
The first relation in (\ref{walg1}) is the abelian Kac-Moody algebra $\u1$, satisfied
by the generators $\w^0_\ell$, and the third one is the Virasoro algebra, obeyed 
by the generators $\w^1_\ell$.
The operators $\wb^i_\ell$ satisfy the same algebra (\ref{walg1}) 
with central charge ${\overline c}=1$ 
and commute with the all the operators $\w^i_\ell$. For this reason, the complete 
Hilbert space
of the CS EFT is a $(c,{\overline c})=(1,1)$ CFT, but since both chiral ($R$) and
antichiral ($L$) sectors are isomorphic, we will often consider one of them for
the sake of simplicity.
The $c=1$ $\winf$ algebra can be realized by either fermionic (more properly, fermion bilinears) or bosonic
operators. In this paper we will focus on the second one (please see \cite{flsz,cfslz}
for further details). 
We remark that the complete factorization of (\ref{hcsf2}) into chiral
and antichiral sectors is the consequence of a Bogoliubov transformation 
that preserves the $\winf$ algebra.
Both sectors are mixed 
by backward scattering terms in the fermionic form of the hamiltonian obtained 
after the thermodynamic limit is taken \cite{flsz,cfslz,boze3}. Moreover, this transformation also performs the bosonization employed here, since the $\winf$ has bosonic and fermionic 
realizations. 
Note that the momentum dependent interactions in (\ref{hcsf2}) are repulsive (attractive)
for $\xi > 1$ ($\xi < 1$). Coupling the system to an external electromagnetic field
induces electric currents by the Laughlin mechanism \cite{laugh}, which implies the quantization of 
$\xi = 1,3,5,\dots$ \cite{cs-reviews}. However, this mechanism implies that the CS particles in (\ref{ham})
are charged, with screened Coulomb interactions (see, {\it i.e.} \cite{boze2}). 

The mode operators (\ref{walg1}) define the $\winf$ currents (fields) $W^i(z)$
on the conformal plane by their Laurent expansion:
\beq
W^i(z)\ \equiv\ \sum_{n=-\infty}^{\infty}\ W^i_n\ z^{-n-i-1}\ .
\label{fourv}
\eeq
These chiral bosonic fields may be represented in terms of a single one, $\varphi(z) \simeq N (z)$ (particle
number field), such that its currents are associated to the $\winf$ operators as $W^{0}(z) \simeq \partial_{z} \varphi(z)$,
$W^{1}(z) \simeq \left ( \partial_{z} \varphi(z) \right)^{2}$ and so on (see, {\it i.e.}, \cite{wref}). 
The $\winf$ currents on the cylinder are found by using
the mapping (\ref{confmap}). For the case of the lowest spin field, 
which is linearly related to the density operator, we have \cite{bz}.
\beq
W^0_R (u)\ = \frac{z}{R} W^0(z)\ .
\label{wcyl}
\eeq
We now discuss the spectrum of (\ref{hcsf2}) in the language of CFT. 
There are both charged and neutral excitations with respect to the fermion number abelian charge. 
In the fermionic description, the $\winf$ operators are usually 
denoted as $\v^i_\ell$  and obey the algebra (\ref{walg1}) with $\xi = 1$. 
The charged highest 
weight states of the $\winf \times {\overline \winf}$ algebra
are obtained by adding $\Delta N$ particles to the fermionic
Fock  ground state, and by moving $\Delta D$ 
particles from the left to the right Fermi point; they are 
denoted by $|\Delta N,\Delta D\rangle_0$.
The neutral particle-hole excitations on top of these are the descendant states, 
$$
|\Delta N , \Delta D ; \{k_i\},\{{\overline k}_j\} 
\rangle_0 \ = 
\v^0_{-k_1} \dots \v^0_{-k_r} \vb^0_{-{\overline k}_1}
\dots \vb^0_{-{\overline k}_s}
|\Delta N , \Delta D \rangle_0~~~~,
$$
with the ordering $k_1 \ge k_2 \ge \dots \ge k_r > 0$, and
${\overline k}_1 \ge {\overline k}_2 \ge \dots 
\ge {\overline k}_s > 0$. 
The number of particle-hole excitations is given by $r$ and $s$. 
The charges associated to these states are
\barr
\v_0^0 ~|\Delta N , \Delta D ; \{k_i\},\{{\overline k}_j\} 
\rangle_0&=& \left(\frac{\Delta N}{2} \ + \Delta D  \right)
|\Delta N , \Delta D ; \{k_i\},\{{\overline k}_j\} \rangle_0~~~,
\nl
\vb_0^0 ~ |\Delta N , \Delta D ; \{k_i\},\{{\overline k}_j\} 
\rangle_0 \ &=& \left(\frac{\Delta N}{2} \ - \Delta D  \right)
|\Delta N , \Delta D ; \{k_i\},\{{\overline k}_j\} \rangle_0~~~.
\label{deltandeltad}
\earr
In terms of the bosonized operators basis $\w^i_\ell$ the highest weight vectors,
$|\Delta N ; \Delta D \rangle_W$, are still
characterized by the numbers $\Delta N$ and $\Delta D$ with the
same meaning as before, but their charges are different.
More precisely
\barr
\w_0^0 ~|\Delta N ; \Delta D \rangle_W  &=&
\left(\sqrt{\xi}\,\frac{\Delta N}{2}+
\frac{\Delta D}{\sqrt{\xi}}\right)
|\Delta N ; \Delta D \rangle_W \nl
\wb_0^0 ~|\Delta N ; \Delta D \rangle_W  &=&
\left(\sqrt{\xi}\,\frac{\Delta N}{2}-
\frac{\Delta D}{\sqrt{\xi}}\right)
|\Delta N ; \Delta D \rangle_W~~~.
\label{vdo}
\earr
The highest weight states $|\Delta N , \Delta D \rangle_W$
together with their descendants, denoted by
$|\Delta N , \Delta D ; \{k_i\},\{{\overline k}_j\} \rangle_W$,
form the bosonic basis for the
theory, that has no simple expression in
terms of the free fermionic degrees of freedom.

The exact energies of these excitations in the bosonic basis are given by:
\barr
{\cal E}&=&
\left\{ \ E_{CS}\
\left[\frac{\sqrt{\xi}}{4}\,Q+\frac{1}{N}
\left(\frac{1}{2}\,Q^2+k\right)
+\frac{1}{N^2}\left(\frac{1}{3\sqrt{\xi}}\,Q^3
-\frac{\sqrt{\xi}}{12}\,Q\right.\right.\right.
\label{eba} \\
&&+\ \left.\left.
\frac{2k}{\sqrt{\xi}}\,Q + \frac{1}{\xi} \sum_{j=1}^{r}k_j^2
-\sum_{j=1}^{r} \left(2j-1\right) k_j\right)\right]
+\left(Q\ \leftrightarrow \ {\overline Q}~,
{}~\{k_j\} \ \leftrightarrow \ \{{\overline k}_j\} \right)\Bigg\}~~~,
\nonumber
\label{exacte}
\earr
where
$$ k \ =\ \sum_{j=1}^{r} k_j ~~~~,~~~~
{\overline k} \ =\ \sum_{j=1}^{\overline r} {\overline k}_j
$$
are the levels, the total dimensionless momenta ({\it i.e.}, physical momenta in units of $1/R$) above the corresponding 
highest weight states. 
The eigenvalues of $\w_0^0$ and $\wb_0^0$, respectively, are
\beq
Q=\sqrt{\xi}\,\frac{\Delta N}{2}+
\frac{\Delta D}{\sqrt{\xi}}~~~~,~~~~
{\overline Q}=\sqrt{\xi}\,\frac{\Delta N}{2}-
\frac{\Delta D}{\sqrt{\xi}}~~~.
\label{Q}
\eeq
Moreover, the integers $k_j$ are ordered
according to $k_1\geq k_2\geq \dots \geq 0$,
and are different from zero only if $r << \sqrt{N}$, and ${\overline r}  << \sqrt{N}$, {\it i.e.}, within the range of validity of the EFT \cite{flsz,clz}.
The number of independent operators (particle-hole number) $P(k)$  at each level $k$ is given by the generating function \cite{bpz}:
\beq
\frac{1}{\prod_{i=1}^\infty \left( 1-q^i \right)\ }\ =\
\sum_{k=0}^{\infty}\ P(k)\ q^{k}\ .
\label{dede0}\eeq
The solitonic charges $Q$ and ${\overline Q}$ are the zero mode charges of a non-chiral
bosonic field (which is the sum of chiral and antichiral bosons) 
compactified on a circle of radius \cite{kaya} :
\beq
r = \frac{1}{\sqrt{\xi}}~~~.
\label{rtil}
\eeq
The partition function of this $c=1$ CFT is known to be invariant 
under the duality symmetry $r \leftrightarrow 1/(2r)$, which in our language 
is equivalent to the mapping between repulsive and attractive regimes $\sqrt{\xi} \leftrightarrow 2/\sqrt{\xi}$.
The action of this mapping on the charges (\ref{Q}) is to interchange 
$\Delta N$ with $\Delta D$, such that $Q$ remains unchanged and ${\overline Q}$
maps onto minus itself. Some known identifications are: $\xi = 2$  
is the self-dual point,  $\xi = 1$  the free fermion point and $\xi = 1/2$ 
the Kosterlitz-Thouless point \cite{bpz}.
Notice that the zero modes are the only links between the $L$ and $R$
Fermi points within the framework of the bosonic EFT.

Summing up, the spectrum (\ref{exacte}) describes both charged and neutral
low lying excitations. The charged (with respect to the $U(1)$ symmetry of fermion number) ones are
labeled by $Q$, which is interpreted as a {\it soliton number}, represented in CFT 
by local vertex operators \cite{bpz}. The 
neutral excitations are labeled by the integers $k_j$ 
and correspond to {\it particle-hole}-like excitations. This analysis originated 
in the fermionic picture extends to the bosonic one \cite{clz}.
The structure of the spectrum is familiar in CFT: the charged excitations are 
highest weight states 
and the neutral fluctuations correspond to the Verma modules on top of each one 
of them \cite{bpz}.
That is to say that the EFT describes a uniform density ground state that may have 
solitons (located lumps or valleys) and fluctuations about them as the complete set of 
low lying excitations. This analysis is in complete analogy with the corresponding one 
for the QHE \cite{laugh}.

The Hilbert space of the EFT of the CS model is factorized in independent chiral and antichiral
sectors, which are isomorphic (due to parity invariance of the model). Without loss of generality, we consider the
chiral sector in the following. 
We first consider the $c=1$ free chiral boson theory, also know as the Tomonaga-Luttinger model.
This theory shares the same Hilbert space as the CS EFT, with
different time evolution of its states.
From the CFT point of view, this theory has a time evolution given by the hamiltonian $H_{0}=v W_{0}^{1}/R$,
where $W_{0}^{1} =L_{0}$, that is, the zero mode of the Virasoro operator \cite{bpz} .
The energy spectrum of the conformal hamiltonian in the basis (\ref{exacte}) is 
$E_{0}(k)=v k/R$, that is, the relativistic rotational energy of a particle with dimensionless angular
momentum $k$. Alternatively, this system may be viewed as one of free chiral phonons. 
For simplicity of notation, we define the ground state  $|\ \Omega\ \rangle\ =\ |\Delta N ; \Delta D \rangle_W$

The DFS is a response function to external density perturbations, and is defined as:
\beq
S\left ( q, \omega \right )\ =\ \frac{1}{4\pi^2}\ \int_{-\infty}^{\infty} dt \int_{0}^{2\pi R} dx\ 
{\rm e}^{ i\left( \omega t -qx \right)}\ \langle\ \Omega\ |\ W_R^{0}\left( u \right) W_R^{0}\left( v \right)\ |\
\Omega\ \rangle\ ,
\label{dfsdef}
\eeq
where $u=\tau v -ix$, $v=0$ and $z=\exp (u/R)$, $w=\exp (v/R)=1$. Here $t$ denotes Minkowski time. 

The strategy is to employ the given hamiltonian to time evolve the relevant field in the Heisenberg picture to 
evaluate the two-point correlation function. This may be done applying the Baker-Campbell-Hausdorff formula.  We have:
\beq
\langle\ \Omega\ |\ W_R^{0}\left( u \right) W_R^{0}\left( v \right) |\ \Omega\ \rangle\ =\ 
\langle\ \Omega\ |\ {\rm e}^{ itH_{0}}\ W_R^{0}\left( -ix \right)\ {\rm e}^{ -itH_{0}}\ W_R^{0}\left( 0 \right)\ 
|\ \Omega\ \rangle\ 
\label{dfs1}
\eeq
We rewrite this expression in terms of the fields in the conformal plane:
\beq
\langle\ \Omega\ | W_R^{0}\left( u \right) W_R^{0}\left( v \right)\ |\ \Omega\ \rangle\ =\ 
\frac{z}{R^2}\ \sum_{n=0}^{\infty}\ \frac{\left ( -i t \right)^n }{n!}\ 
\langle\ \Omega\ |\ {}_n [\ W^{0}\left( z\right) , H_{0}\ ]\ W^{0}\left( 1\right)\ |\ \Omega\ \rangle\ ,
\label{dfs2}
\eeq
with $z=\exp(-i\theta )$. Here 
\beq
{}_n [\ W^{0}\left( z\right) , H_{0}\ ]\ =\ W^{0}\left( z\right) +\ [\ W^{0}\left( z\right) , H_{0}\ ]
+\ [\ [\ W^{0}\left( z\right) , H_{0}\ ], H_{0}\ ]\ +\ \dots
\label{dfs3}
\eeq
From this expansion it is clear that only the neutral ($Q=0$) states in the Hilbert space may contribute to
(\ref{dfsdef}). Given that the expectation value of the two-point correlator is evaluated in the ground
state, we focus on the Verma module of the vacuum. That is, the space of all neutral excitation with 
a given integer momentum $k=1,2,\dots$ and an integer number $n_{ph}=1,2,\dots$ of $W_{-k}^0 $ operators 
applied to the ground state, which may also be viewed as the number of phonons on top of it.

Equation (\ref{dfs2}) may be evaluated to each order in $n$ by inserting projectors on the relevant 
sectors of the Hilbert space. 
We first consider the (neutral)  $n_{ph} =1$ excitations with dimensionless integer momentum $k$ of the form 
\beq
|\ k\ \rangle\ =\ W_{-k}^0 |\ \Omega\ \rangle\ ,\ k=1,2,\dots \ .
\label{wok}
\eeq
These excitations are bosonic, but get their name from the analogous $\winf$ fermionic operator. 
The relevant projector onto the subspace of these excitations is defined as:
\beq
\mathcal P_{1}\ =\ \sum_{k=1}^{\infty}\ \frac{1}{\xi k }\ |\ k\ \rangle\ \langle\ k\ |\ .
\label{proj}
\eeq
It is easy to verify that this is indeed a projector. 
While evaluating the different terms in the expansion (\ref{dfs2}) after insertion of $\mathcal P_{1}$,
some matrix elements are useful. For example, 
\beq
\langle\ k\ |\ W^{0}\left( z\right) |\ \Omega\ \rangle\ =\ \xi k z^{k-1}\ ,
\label{matrixele1}
\eeq
which are easily obtained from the (def of W) and (conmutators). 
Taking the adjoint of this relation we obtain \cite{bpz}, as well:
\beq
\langle\ \Omega\ |\ W^{0}\left( z\right) |\ k\  \rangle\ =
\left(
\langle\ k\ |\ W^{0}\left( z\right) | \Omega\ \rangle\ 
\right)^{\dagger}\ = \xi k \left( z^{-1} \right)^{k-1} \left( z^{-1} \right)^{2}\ .
\label{matrixele2}
\eeq
Other relevant matrix elements involve the hamiltonian, for example: 
\barr
\langle\ k'\ |\ \frac{v}{R} W_{0}^{1}\ |\ k\ \rangle &=& \frac{v}{R} \xi k^{2} \delta_{k,k'} \nl
\langle\ k'\ |\ \left( \frac{v}{R} W_{0}^{1} \right)^{2} |\ k\ \rangle 
&=& \left( \frac{v}{R} \right)^{2} \xi k^{3} \delta_{k,k'}\ ,
\label{h0matrix}
\earr
with straightforward generalizations.

Before proceeding with the computation of (\ref{dfsdef}) we note that subspaces of the neutral
sector of excitations with $n_{ph}=2,3\dots$ (defined by projectors generalizing (\ref{proj})) give zero 
contributions to \ref{dfsdef}) for the relevant matrix 
elements vanish:
\beq
\langle\ k_1,k_2,\dots\ |\ W^{0}\left( z\right) |\ \Omega\ \rangle\ =\ 0\ .
\label{matrixelen}
\eeq
This means that only the subspace defined by (\ref{wok}) is relevant for the computation of (\ref{dfs2}).

Therefore, the first terms of the expansion (\ref{dfs2}), are:
\barr
&&\langle\ \Omega\ |\ W_R^{0}\left( u \right) W_R^{0}\left( v \right)\ |\ \Omega\ \rangle\ = 
\frac{\xi}{R^2}\ \left[ \
\sum_{k=1}^{\infty} k\ z^{-k}\ +\ \left( -\frac{itv}{R} \right)\ \sum_{k=1}^{\infty} k^{2}\ z^{-k}\ + \right. \nl
&&+\ \left.\frac{1}{2!} \left( -\frac{itv}{R} \right)^{2}\ \sum_{k=1}^{\infty} k^{3}\ z^{-k}\ +\ \dots 
\right] 
\label{dfs4}
\earr
Replacing $z=\exp(-ix/R)$ in (\ref{dfs4}) we may compute the DSF (\ref{dfsdef})
for the free theory, $S_{0}$:
\barr
&&S_{0}\left ( q, \omega \right )\ =\ \frac{1}{4\pi^2}\ \int_{-\infty}^{\infty} dt \int_{0}^{2\pi R} dx\ 
{\rm e}^{ i\left( \omega t -qx \right)}
\frac{\xi}{R^2}\ \left[ \
\sum_{k=1}^{\infty} k\ {\rm e}^{ ikx/R}\  \right. \nl
&&+\ \left.
\left( -\frac{itv}{R} \right)\ \sum_{k=1}^{\infty} k^{2}\ {\rm e}^{ ikx/R}\ +\
\frac{1}{2!} \left( -\frac{itv}{R} \right)^{2}\ \sum_{k=1}^{\infty} k^{3}\ {\rm e}^{ ikx/R}\ +\ \dots 
\right]\ .
\label{dfs5}
\earr
Hence
\barr
&&S_{0}\left ( q, \omega \right )\ =\ \frac{\xi}{2\pi R}\ \int_{-\infty}^{\infty} dt 
{\rm e}^{ i \omega t }\
\left[ \
\sum_{k=1}^{\infty} k\ 2\pi \delta_{qR,k}\ +\ \left( -\frac{itv}{R} \right)\ 
\sum_{k=1}^{\infty} k^{2}\ 2\pi \delta_{qR,k}\ + \right. \nl
&&+\ \left.\frac{1}{2!} \left( -\frac{itv}{R} \right)^{2}\ \sum_{k=1}^{\infty} k^{3}\ 2\pi \delta_{qR,k}\ +\ \dots 
\right] \nl
&&=\ \frac{\xi}{R}\ \int_{-\infty}^{\infty} dt 
{\rm e}^{ i\omega t }\
\left[ \
qR\   +\ \left( -\frac{itv}{R} \right)\ (qR)^{2}\ +\ 
\frac{1}{2!} \left( -\frac{itv}{R} \right)^{2}\
(qR)^{3}\  +\ \dots 
\right] \nl
&&= {\xi q}\ \int_{-\infty}^{\infty} dt 
{\rm e}^{ i( \omega  -vq)t }\ =\ {\xi q} \delta\left(  \omega  -vq\right)
\label{dfs7}
\earr
This is a well known result for the Tomonaga-Luttinger liquid \cite{1dliquidsrev,markhof,teber}. It
means that the response of the free boson system to an external perturbation of the density
takes place in resonance with the eigenfrequency  $\omega  = vq $  only, in correspondence to the
uniform circular motion around the spatial coordinate with constant wave velocity $v = \omega /q$. 
The fact that this response is sharp means that the eigenstate is a true one, stable in time.

We remark that considering both chiralities yields a global
response function that is the even extension of (\ref{dfs7})
in $\omega$, associating negative frequencies to
angular motion in the opposite sense. Formally, it corresponds to 
changing the sense of the plasma velocity, $v \to - v$, in  $E_0$, keeping the momenta ${\overline k}$ positive. 
This amounts to 
global momentum conservation on the entire isolated system with both chiralities. 

We proceed with the same scheme in the case of the CS hamiltonian (\ref{hcsf2}).
As in the previous case, we consider the the Hilbert space of neutral excitations of the chiral sector only, 
implying that all states satisfy $Q=0$. This means that the hamiltonian restricted 
to this subspace does not involve the operators $W_0^0 $. 
The main difference in the computation of the DSF relies in the different values of the matrix elements 
of the CS hamiltonian:
\barr
&&\langle\ k'\ | 
\frac{E_{CS}}{N}
\left[
\w_0^1 +
\frac{1}{N}\left(\frac{1}{\sqrt{\xi}}\,\w_0^2
-\frac{( \xi -1)}{\xi}\,\sum_{\ell=1}^\infty
\,\ell~\w_{-\ell}^0\,\w_\ell^0\right)
\right]
|\ k\ \rangle\ 
= \nl
&&\qquad\qquad \frac{E_{CS}}{N} \xi\ \left[ k^{2}\ -\ \frac{(\xi -1)}{N}\ 
k^{3}\ \right]
\delta_{k,k'} 
\label{hcsmatrix1}
\earr
\barr
&&\langle\ k'\ |\ 
\left(\frac{E_{CS}}{N}\right)^{2}\ 
\left[
\w_0^1 +
\frac{1}{N}\left(\frac{1}{\sqrt{\xi}}\,\w_0^2
-\frac{( \xi -1)}{\xi}\,\sum_{\ell=1}^\infty
\,\ell~\w_{-\ell}^0\,\w_\ell^0\right)
\right]^2
|\ k\ \rangle\ = \nl
&&\qquad\qquad \left(\frac{E_{CS}}{N}\right)^{2}\  \frac{\xi}{k}\ \left[ k^{2}\ 
-\ \frac{(\xi -1)}{N}\ k^{3}\ \right]^{2} \delta_{k,k'} 
\label{hcsmatrix2}
\earr
Generalizations are, again, straightforward. 

We therefore obtain the following expression for the two-point density
correlator evolved with the CS hamiltonian:
Therefore, the first terms of the expansion (\ref{dfs2}), are:
\barr
&&\langle\ \Omega\ |\ W_R^{0}\left( u \right) W_R^{0}\left( v \right) |\ \Omega\ \rangle = 
\frac{\xi}{R^2}\ \left[ \
\sum_{k=1}^{\infty} z^{-k}k \right.\nl 
&&+\ \left(-it\frac{E_{CS}}{N} \right)\ \sum_{k=1}^{\infty} z^{-k} 
\left( k^{2}\ -\ \frac{(\xi -1)}{N}\ k^{3}\ \right) \nl
&&+\ \left.\frac{1}{2!} \left(-it\frac{E_{CS}}{N}\right)^{2}\ \sum_{k=1}^{\infty} z^{-k}
\frac{1}{k}\ \left( k^{2}\ -\ \frac{(\xi -1)}{N}\ k^{3}\ \right)^{2}
 +\ \dots 
\right] 
\label{dfscs1}
\earr
From this expression, it is straightforward to find the DSF of the CS model, $S_{CS}$
which is given by:
\beq
S_{CS}\left ( q, \omega \right )\ =\ {\xi q} \delta\left[  \omega\ -\ \frac{E_{CS}}{N}
\left( qR\ -\ \frac{(\xi -1)}{N}\ {qR}^{2}\ \right) \right]
\label{dsfcs}
\eeq
This is the main result of our paper. We find again that the response of the system to a density perturbation
takes place at one frequency only. Noticing that 
$E_{CS} /N =v/R$, we find that the corresponding eigenfrequency may be expressed as
\beq
\omega\ =\ \omega_{CS}\ =\ \frac{vq}{R}
\left[ 1\ -\ \frac{(\xi -1)}{N}\ {qR}\ \right]\ .
\label{omegacs}
\eeq
The relativistic rotational energy of a particle with dimensionless angular
momentum $k$ is now slightly smaller than in the free case due to 
interactions. A picture based on interacting phonons may be also considered.
Equivalently, the self interactions of the bosonic field tend to slow down the angular
velocity,{\it i.e.}, making the rotational period longer.  
Among the assumptions made in the definition of the CS EFT are the conditions $N >> 1$, $k=qR \leq \sqrt{N}$
and $\xi \geq 1$. Moreover, we consider the range of couplings $1 \leq \xi << \sqrt{N}$, 
leaving away here the feasibility of making sense of the EFT for larger couplings ($\sqrt{N}$ is 
the natural UV cutoff of the EFT \cite{clz}).
Under these conditions, we have that the frequency 
shift with respect to the free theory is very small, that is, writing $\omega
= \omega_0 + \Delta \omega$, with $\omega_0 = vq/R$ and 
$\Delta \omega = -\ \frac{(\xi -1)}{N}\ (vq /R){qR}\ $, with $|\Delta  \omega| << \omega_0 $.
Defining the rotational period
as $T=2\pi /\omega$ we get $T \simeq T_{0} + \Delta T$, with $T_{0}=2\pi /\omega_{0}$ being the free
period, and $\Delta T /T_{0} = -\Delta \omega /\omega_{0}$. Moreover, 
$\Delta \omega (q) = -\ \frac{(\xi -1)}{N}\ vq^{2}\ $, which implies, in coordinate space 
$\Delta \omega (x) =\ \frac{(\xi -1)}{N}\ v 2\pi\delta^{(2)} (x)$, which
is negative due to the concavity of the second derivative of the Dirac delta function.
The physical picture that emerges is that the CS interaction within the bosonic fields has the effect
of creating defects arbitrarily located in the spatial circumference that slow down the otherwise free 
propagating waves. 
In the hydrodynamic picture, the motion of the fluid is described by the Benjamin-Ono equation,
with non linear dispersion relation (\ref{omegacs}).
This picture, however cannot be complete by considering 
one chirality of the EFT only. A real effective potential
as the one considered above cannot couple to a chiral current.
Therefore, the picture based on the defect potential (trap) 
makes sense only when both chiral and antichiral components 
are considered (a similar discussion regarding the coupling 
of the EFT to an external electromagnetic field is given in 
\cite{boze2}). Similarly, the orbital angular momentum 
with respect to the center of the spatial circumference
would not be conserved by considering one chiral component 
only. Considering both yields global orbital angular momentum
conservation of an isolated CS system. 
The remark on the frequency response of the DSF regarding the consideration
of both chiralities in the free case holds here as well.

The DSF for the CS model has been computed in its first quantized formulation in \cite{pustilink}. 
We now compare the result obtained with the EFT to it. The form of the DSF in \cite{pustilink}
is approximately a narrow rectangle of width $\delta \omega = (\xi +1 )q^{2}/{2m}$, height $q/\delta \omega $ and 
special behavior at the extremes of the support, $\omega_{+} = \omega_{0} + q^{2}/{2m}$ and
$\omega_{-} = \omega_{0} -\xi q^{2}/{2m}$ and $\omega_{0} =vq$ (with the plasma velocity $u=v$ (\ref{vti2})).
The mass parameter $m$ corresponds to $N/2$ in our formulation ($N \to \infty$). The processes considered 
in \cite{pustilink} include fermionic transitions of high energy, like those of backward
scattering. These interactions have been integrated out while obtaining the bosonic formulation of EFT
of the CS model, so we do not expect these processes to be present in the response function of the
EFT. These high energy fermionic processes are also considered responsible of the emergence 
of the bosonic ground state \cite{boze3}. Moreover, given that we focus in this paper in the repulsive
regime, of the two branches considered in \cite{pustilink}, only $\omega_{+}$ plays a relevant role here.
Therefore, we expect that the results of \cite{pustilink} when $m \to \infty$,
may be confronted to (\ref{dsfcs}). In the limit $\delta \omega \to 0 $, the rectangle response function in \cite{pustilink} 
is expected to converge to a Dirac delta function, in analogy with 
the response function of the free boson theory (\ref{dfs7}).
Here  we provide a more detailed derivation of this result. Firstly, notice that $\omega_{+}$ 
displays the same functional structure as $\omega_{CS}$ (\ref{omegacs}), and therefore
play similar roles in both first and second quantized formulations. Consider the approximate 
expression of $S(q,\omega)$ found in \cite{pustilink}:
\beq
S\left ( q, \omega \right )\ \simeq\ 
\left | \frac{\omega_{+}- \omega}{\delta \omega}\right |^{\xi -1}\ \frac{m}{q}\ ,
\label{dfscspus}
\eeq
valid for the region $0 < \omega_{+}- \omega << \delta \omega$, with possibly omitted prefactors.
For the purposes of comparison we have use the notation $\xi $ for the coupling constant. Fourier
transforming into time domain, we get:
\beq
S\left ( q, t \right )\ \simeq\ \frac{m}{q}\ \left ( \frac{1}{\delta \omega}\right )^{\xi -1}\
\int_{\omega_{+} - \delta \omega}^{\omega_{+}}\ d\omega\ 
{\rm e}^{ -i\omega t}\ 
\left ( {\omega_{+}- \omega} \right )^{\xi -1}\ \
\label{dfscspustim}
\eeq
Evaluating this expression in the limit $\delta \omega t << 1$, we obtain:
\beq
S\left ( q, t \right )\ \simeq\ \frac{m}{q}\ \delta \omega\ 
{\rm e}^{ -i\omega_{+} t}\ =\ 
\frac{(\xi +1 )}{2}\ q\ {\rm e}^{ -i\omega_{+} t}\ ,
\label{dfscspustim2}
\eeq
which corresponds, up a proportionality factor independent of $q$, to (\ref{dsfcs}),
with $\omega_{+}$ playing the role of $\omega_{CS}$.

Furthermore, the first frequency moment of the DSF is proportional to $q^2$ 
for systems with momentum independent interactions,  and this result is known as the $f$-sum rule 
\cite{pines,fsumvio} . We verify that it holds for the free boson theory but 
not for the CS EFT, with hamiltonian given by (\ref{hcsf2}):
\barr
&&\int_{0}^{\infty}d\omega\ \omega\ S_{0}\left ( q, \omega \right )\ =\ \frac{\xi v}{R}\ q^{2}\ \nl
&&\int_{0}^{\infty}d\omega\ \omega\ S_{CS}\left ( q, \omega \right )\ =\ \frac{\xi v}{R}\ q^{2}\ 
\left [ 1-\frac{(\xi -1)}{N} (qR) \right ]
\label{fsumrules}
\earr

In conclusion, a detailed comparison between the DSFs for the EFT of the CS model and that corresponding 
to the first quantized formulation of it
was provided,  
and complete agreement in the parameter 
region in which both
results overlap was found. 
Further support to the equivalence between first and second quantized
treatments of the Calogero-Sutherland model
was, therefore, provided. 
The advantages of integrability in the first formulation 
are mirrored by those from the algebraic structure of the second. 
In both cases, calculability by analytic methods
is a major attractive feature, that allows for a deeper understanding of the physical properties of the system.
The long time physical picture that emerges for one chiral component of the CS EFT is that of a weakly nonlinear 
quantum fluid in one dimension, 
with waves propagating on the circular spatial
geometry. In the repulsive regime, periods are slightly slower due to the interactions, that may be represented 
as kicks that oppose 
the direction of motion, localized in the circumference . 

%-------------------------------------------------------------------
%
\def\NP{{\it Nucl. Phys.\ }}
\def\PRL{{\it Phys. Rev. Lett.\ }}
\def\PL{{\it Phys. Lett.\ }}
\def\PR{{\it Phys. Rev.\ }}
\def\IJMP{{\it Int. J. Mod. Phys.\ }}
\def\MPL{{\it Mod. Phys. Lett.\ }}

\end{document}